\documentclass[aps,pra,twocolumn,showpacs,superscriptaddress]{revtex4-1}
\usepackage{amsmath}
\usepackage{amssymb}
\usepackage{mathtools}
\usepackage{graphicx}
\renewcommand{\vec}[1]{\mathbf{#1}}

\begin{document}
\title{Itinerant Ferromagnetism in SU($N$)-Symmetric Fermi Gases at Finite Temperature: First Order Phase Transitions and  Time-Reversal Symmetry}

\author{Chen-How Huang }
 \affiliation{Donostia International Physics Center (DIPC), 20018 Donostia-San Sebastian, Spain}
\author{Miguel A. Cazalilla }
  \affiliation{Donostia International Physics Center (DIPC), 20018 Donostia-San Sebastian, Spain}
\affiliation{Ikerbasque, Basque Foundation for Science, 48013 Bilbao, Spain}
\date{\today}
\begin{abstract}
  At  temperatures well below the Fermi temperature $T_F$, the coupling of magnetic fluctuations  to  particle-hole excitations in a two-component Fermi gas makes the   transition to itinerant ferromagnetism a first order phase transition. This effect is not described by the paradigm of Landau's  theory of phase transitions, which  assumes  the free energy is an analytic function of the order parameter and  predicts a second order phase transition. On the other hand, despite that larger symmetry often introduces larger degeneracies in the low-lying states, here we show that for a Fermi gas with SU($N > 2$)-symmetry in three space dimensions the ferromangetic phase transition is first order in agreement with the predictions of Landau's theory [M. A. Cazalilla \emph{et al}. New J. of Phys. {\bf 11} 103033 (2009)]. By performing unrestricted Hartree-Fock calculations for an SU($N > 2$)-symmetric Fermi gas with short range  interactions, we find the order parameter undergoes a finite jump across the transition. In addition, we do not observe any   tri-critical point up to temperatures $T \simeq 0.5\: T_F$, for which the thermal smearing of the Fermi surface is  subtantial. Going beyond mean-field, we find that the coupling of magnetic fluctuations to  particle-hole excitations  makes the transition  more abrupt and further enhances  the tendency of the gas to become fully polarized for smaller values of $N$ and the gas parameter $k_F a_s$. In our study,  we  also clarify the role of time reversal symmetry in the microscopic Hamiltonian and obtain the temperature dependence of Tan's contact. For the latter,  the presence of the tri-critical point for $N = 2$ leads to a more pronounced temperature dependence  around the transition than for SU($N > 2$)-symmetric gases.
\end{abstract}
\maketitle

\section{Introduction}
Itinerant ferromagnetism (FM) appears to be a rather common and robust phenomenon
in the realm of metals. Yet,
a complete microscopic understanding  is still lacking despite  much theoretical effort being devoted to its study.  Bloch  made an early  attempt to explain it in 1929~\cite{Bloch_1929}:  Using  what later became known as the Hartree-Fock (HF) approximation, he found that the Coulomb interaction in an electron gas would cause a phase transition to a spin polarized state at low densities. 
This result was criticized by Wigner, who argued that the HF approximation neglects  important electron correlation effects that, at low-densities, suppress the FM in favor of crystaline order~\cite{Wigner_PhysRev.46.1002}. 

 However,  electron density is not small in most metals and  the long-range Coulomb interaction is strongly screened. A more appropriate starting point would be a model with short-range interactions. Indeed, in the 1960's Gutzwiller~\cite{Gutzwiller_PhysRevLett.10.159}, Hubbard~\cite{Hubbard_1963},
 and Kanamori~\cite{Kanamori_1963} motivated by similar concerns about the role of electron correlation effects in ferromagnetism in metals proposed a model currently known as the Hubbard model.
 This model is characterized by two energy scales: $t$, which parametrizes the kinetic energy (i.e. the band-width), and $U$, which parametrizes the strength of the (short-range) interactions.
 Applying the  HF method to this model yields the following condition for itinerant FM:
 \begin{equation}
 U N(E_F) \geq 1. \label{eq:stoner}
 \end{equation}
where $N(E_F)\sim 1/t$ is the density of states at the Fermi energy. This condition is known  as Stoner's criterion and pre-dates the Hubbard model. However,
Kanamori~\cite{Kanamori_1963} argued that   Eq.~\eqref{eq:stoner}  overestimates the tendencies towards itinerant FM. Instead, the renormalization of the Hubbard interaction caused by repeated scattering of quasi-particles makes its occurrence more difficult, and essentially impossible at low densities~\cite{Kanamori_1963}. This expectation has been numerically confirmed by a recent quantum Montecarlo calculation of the ground state energy and spin polarization in the low-density limit of the Hubbard model~\cite{Ceperley_PhysRevA.82.061603}. 

Nevertheless, in the absence of a lattice (i.e. the continuum limit), variational Montecarlo calculations  found that the FM transition takes place~\cite{Pilati_PhysRevLett.105.030405_2010} for values of the gas parameter $k_F a_s \approx 1$, where $k_F$ is the Fermi wave number and $a_s$ the s-wave scattering length
(indeed, the actual value depends on  details of the inter-particle potential beyond the scattering length~\cite{Pilati_PhysRevLett.105.030405_2010}). The value of the transition point is roughly in agreement with the prediction of Stoner's criterion for the Fermi gas. Furthermore, Stoner's model and its more sophisticated version, Landau's theory of phase transitions, predict a continuous phase transition to the itinerant ferromagnet in three spatial dimensions. On the other hand, recent quantum Montecarlo calculations~\cite{Conduit_PhysRevLett.103.207201_2009,Pilati_PhysRevLett.105.030405_2010} and earlier theoretical studies~\cite{Belitz_Kirkpatrick_PhysRevLett.94.247205,Belitz_PhysRevLett.89.247202_2002,Duine_PhysRevLett.95.230403_2005,Conduit_PhysRevA.79.053606_2009}  
found a first order transition. The change of order is a consequence of  the coupling of (ferro-) magnetic fluctuations to particle-hole excitations~\cite{Belitz_Kirkpatrick_PhysRevLett.94.247205,Belitz_PhysRevLett.89.247202_2002,Duine_PhysRevLett.95.230403_2005,Conduit_PhysRevA.79.053606_2009}, a phenomenon akin to the Weinberg-Coleman mechanism of high-energy Physics~\cite{WeinbergColeman_PhysRevD.7.1888}. This mechanism  is suppressed by thermal fluctuations. Thus, above a certain (tri-critical point) temperature, the transition becomes continuous again. This is in agreement the experimental observations of a continuous phase transition in FM metals like Fe and Ni, for which the Curie temperature $T_{c}\approx 1000$K.  
 
Many of the above results concern two-component Fermi gases and by comparison, much less is known about itinerant FM in Fermi systems with higher number of components. In lattice systems, higher number of components can arise from a multi-orbital description of the electronic properties of a  material. In such systems a new energy scale emerges, namely  Hund's coupling, which  favors spin polarization (just like it does in atoms and it is encoded in Hund's rules).
Multi-orbital Kanamori-Hubbard models have been studied using a wealth of techniques, most notably the dynamical mean-field theory (see e.g. Ref.~\cite{georges2013} and references therein). 

In the continuum limit, there are fewer available results  for itinerant FM. 
For $N$ component Fermi gases, it was found in Ref.~\cite{Cazalilla_NJOP_11_2009} using a Landau free energy derived by integrating out the fermionic degrees of freedom of an interacting Fermi gas  with SU$(N)$ symmetry~\cite{Cazalilla_NJOP_11_2009,
Gorshkov_2010,Cazalilla_2014}  the transition to the itinerant ferromagnet is generically first order in three space dimensions. This may appear striking because  higher symmetry is expected to lead to higher degeneracies in the low-lying states, thus 
enhancing quantum criticality. However, as
discussed below, the structure of the FM order parameter in SU($N > 2$) leads to a different form of the Landau free-energy. This difference sets the  SU$(N = 2)$ and SU$(N >2)$ symmetric Fermi gases clearly apart. 

 Recently, Pera and coworkers have carried out HF, 2nd~\cite{Pera1} and 3rd order~\cite{Pera2} calculations at zero temperature for the SU($N$)-symmetric Fermi gases and confirmed the existence of the first order phase transition. Their approach 
assumes a particular pattern of symmetry breaking of SU($N$)  which reduces the ground state energy to a function of a single parameter. The energy is 
then minimized as a function 
of this  parameter. 
To 2nd order in the gas parameter $k_F a_S$ the ground state energy is known analytically from the work of Kanno~\cite{Kanno_PTP_1970}.
Accounting for the
3rd order  goes beyond universality and requires numerical  integration~\cite{Pera2}. 

In this work, we study the trans\textbf{}ition to the itinerant FM at finite temperatures. Our approach is fully numerical and does not assume any particular pattern of SU($N$) symmetry breaking. Indeed,  having $N-1$ diagonal generators and depending on the microscopic details of the model, SU($N$) symmetry may be broken in several different patterns. For a three-dimensional Fermi gas with Dirac delta-interactions, we obtain below  the pattern of symmetry breaking without any \emph{a priori} assumptions by performing unconstrained minimization of the free energy. Thus we find that, unlike the SU($N = 2$)-symmetric interacting Fermi gas,  in the  SU$(N > 2)$-symmetric Fermi gas there is no evidence of a tri-critical point up to $T = 0.5 \: T_F$, even when the coupling of the fluctuations of the magnetization is accounted for. We also report results for the temperature dependence of Tan's contact~\cite{TAN_AnnPhys_2008}, which controls the asymptotic behavior of the momentum distribution. We find that the existence of a tri-critical point leads to a stronger temperature dependence of the contact for SU($N=2$)-symmetric Fermi gases  compared to SU($N > 2$) systems.

The rest of this article is organized as follows: In section~\ref{sec:landau}, we review the most important predictions of Landau's theory of phase transitions for itinerant ferromagnets and discuss the role played by time-reversal symmetry.  In section~\ref{sec:uhf},  we introduce the microscopic model and report our results using the unrestricted Hartree-Fock method at finite temperature.
In section~\ref{sec:2nd_p} we discuss the corrections introduced by an unrestricted minimization of the 2nd order free energy.  In section~\ref{sec:tan}, the temperature dependence of Tan's contact for  $N=2$ and $N >2$ is discusssed. Finally, in section~\ref{sec:conclusion} we provide a discussion of the significance of our results in a broader framework and offer our conclusions. The details of the calculations and methods employed in this work are provided in the Appendices.

\section{Landau Mean-field  Theory and Time-Reversal SYMMETRY} \label{sec:landau}
Let us recall the main results of Landau's theory of phase transitions and clarify  the role played by time-reversal symmetry (TRS) in description of the transition between the paramagnetic gas and the itinerant ferromagnet. 
Independently of the  form of the microscopic Hamiltonian, for an interacting
Fermi liquid with SU$(N)$-symmetry the order parameter of the transition from a paramagnetic gas to an itinerant ferromagnet is the traceless part of the Landau quasi-particle distribution density matrix,
$n_{\beta}^{\alpha}(\vec{p})$~\cite{Cazalilla_NJOP_11_2009}:
\begin{equation}
\bar{M}_{\beta}^{\alpha} = \sum_{\vec{p}}
\left[ n^{\alpha}_{\beta}(\vec{p}) - \frac{\delta^{\alpha}_{\beta}}{N}  n^{\gamma}_{\gamma}(\vec{p}) \right].
\end{equation}
As a rank-2 tensor, $\bar{M}$ belongs to the adjoint representation of  SU$(N)$, and therefore can be expressed as a linear combination of the generators of the Lie algebra:
\begin{equation}
\bar{M} = \sum_{a=1}^{N^2_c-1} m_a T^a, 
\label{eq:groupexp}
\end{equation}
where the Lie algebra generators are normailzed as follows:
\begin{equation}
\text{Tr}\, T^a T^b = \tfrac{1}{2} \delta^{ab}.
\label{eq:norm}
\end{equation}
In general, being $\bar{M}$ a hermitian matrix, it can be diagonalized and the  the free energy  written in terms of its $N-1$ real eigenvalues or, equivalently, in terms of the expectation value of the diagonal generators of SU($N$) (the so-called Cartan subalgebra). In 
 certain microscopic systems like the continuum Fermi gas described below  (cf. Eq.~\eqref{eq:H}),
 the energy is minimized by having the order parameter $\bar{M}$ being proportional to \emph{only one} element of the Cartan (normalized as in Eq.~\ref{eq:norm}). This corresponds  to  a pattern of   breaking the SU$(N)$ symmetry such that SU($N$) $\to$ SU$(N-1)$ $\times$ U($1$). Thus, to describe the broken symmetry phase it is sufficient to use a single scalar (see next section). However, here we want to keep the discussion as general as possible and in the following we will write the Landau free energy in its SU($N$) invariant form. 

  Turning our attention to the actual form of the   of  Landau's  free energy and assuming that in the neighborhood of the phase transition  the matrix elements of $\bar{M}$ are small (or, in the  representation of Eq.~\ref{eq:groupexp},  $|m_a| \ll 1$), the change in free energy  follows from symmetry  and can be written as an analytic series  expansion in terms of the scalar invariants of $\bar{M}$, which, for the SU($N$)-group, takes the form~\cite{Cazalilla_NJOP_11_2009}:
\begin{align}
F-F_0 = \frac{c_2}{2}\text{Tr}\, \bar{M}^2+\frac{c_3}{3}\text{Tr} \, \bar{M}^3+\frac{c_4}{3}\text{Tr}\, \bar{M}^4 +\cdots \label{eq:landau}
\end{align}
Here $c_2$ changes sign for a certain value of the interaction strength and the temperature $T$ determined by Stoner's criterion~\cite{Cazalilla_NJOP_11_2009}. For stability, the value of  $c_4$ is taken to be positive (this is indeed case of the system described by Eq.~\ref{eq:H}). 
Thus, in general, the presence of a cubic term 
implies a first order phase transition~\cite{Cazalilla_NJOP_11_2009}. 
However, the case $N = 2$ is 
special because  the term $\propto\text{Tr}\, \bar{M}^3$ vanishes  along with all terms of the form $\propto\text{Tr}\, \bar{M}^n$ where $n$ is odd. Therefore, Landau's theory predicts a continuous phase transition for SU($N=2$).  

Mathematically, the presence of the cubic term in the Landau free energy  can be traced back to the existence of a second set of symmetric structure constants in the SU($N > 2$) groups which, together with the anti-symmetric structure constants, fully  determine the product of any pair of Lie-algebra generators:
\begin{align}
T^aT^b=\frac{1}{2N}\delta^{ab}+\frac{1}{2}\sum_{c=1}^{N^2-1}\left(d^{ab}_{c}+i f^{ab}_{c}\right)T^c,
\end{align}
Here  $d^{ab}_{c}$ and $f^{ab}_{c}$ are, respectively, fully symmetric and anti-symmetric in the (Roman) indices $a,b$ ($a,b,c=1,\ldots, N^2-1$). For SU($N=2$),   $d^{ab}_{c}=0$  but $d^{ab}_{c}\neq0$ for $N > 2$, which leads to the non vanishing term $\propto \text{Tr} \: \bar{M}^3$ discussed above.

 Finally, let us discuss the role played by the time-reversal symmetry of the Hamiltonian. It can be argued that the presence of the term  $\propto\text{Tr}\, \bar{M}^3$ can be ruled out by supplementing the SU$(N)$ symmetry with TRS. However, as we show in what follows, this  only true for SU($N =2$) and does not hold for   $N >2$.  First of all, let us examine how TRS is implemented in  a Fermi gas with $N > 2$ components. To this end, we first need to determine whether the microscopic Hamiltonian is invariant under TRS. Indeed, this is a non-trivial question for the following reason: Let us first recall that under a time-reversal transformation described by the anti-unitary operator $\mathcal{T}$ the total spin operator $\vec{F}$ changes as $\mathcal{T}^{-1} \vec{F}\mathcal{T} = -\vec{F}$. Thus, for systems whose constituent particles are half-integer spin fermions,  the Hamiltonian is invariant under TRS provided it can be expressed in terms of quantum fields that are Kramers pairs. This is not the case if, for example, a $N=4$ component atomic gas sample is prepared as a mixture of $\{-7/2,-5/2,-3/2, -1/2\}$ $^{173}$Yb nuclear spin components (or any other combination of an even number components such that the sum of the $F_z$ projection of the components is not zero). And it is certainly never the case for $N$ 
 odd because there is always at least one component that lacks a Kramers pair. Note that, in all those cases, the Hamiltonian still exhibits SU$(N)$ symmetry as the interactions are very insensitive to the nuclear spin orientation of the atoms in the mixture. However, the \emph{physical} TRS is broken explicitly by the choice of the systme constituents. This situation does not arise in solid state Physics, but the capability of trapping  mixtures of different nuclear-spin components of alkaline-earth cases makes it possible in atomic Physics~\footnote{Alternatively, we can imagine starting with a Hamiltonian for $N^{\prime}$ components that displays TRS and 
 forcing some of the nuclear spin components out of the system by making the value of $\mu_{\alpha}$ for them very negative. This is 
 equivalent to applying very strong time reversal symmetry breaking ``magnetic'' fields which in addition reduce the symmetry from SU($N^{\prime}$) to SU($N < N^{\prime}$).}. 

Having  established the conditions for the microscopic Hamiltonian to display TRS, let us next discuss how the order of the ferromagnetic transition is affected by it. To this end, we fist explain the transformation properties of the magnetization matrix $\bar{M}$ under time-reversal. The full details are provided in the Appendix, where it is shown that  under $\mathcal{T}$ the order parameter $\bar{M}$ transforms  as follows:
\begin{equation}
\bar{M} \to \Lambda^{T} \bar{M}^{T
} \Lambda, \label{eq:trs}
\end{equation}
where (for $N = 2F+1$ and half-integer $F=\tfrac{1}{2},\tfrac{3}{2},\frac{5}{2},
\ldots$ )
\begin{equation}
\Lambda =     \left( 
\begin{array}{ccccc}
0 & 0 & \cdots & 0 & -1\\
0 & 0 & \cdots &+1 & 0\\
\vdots & \vdots & \cdots &  \vdots &\vdots  \\
0 & -1 & \cdots & 0 & 0\\
+1 & 0 & \cdots &0 & 0
\end{array}
\right). \label{eq:lambda}
\end{equation}
The matrix $\Lambda$
acts on the SU($N$) spinor quantum field $\Psi^T(\vec{r}) = \left(c^{1}(\vec{r}), \ldots, c^N(\vec{r}) \right)$ where the fields are arranged  in decreasing value of  $F^z$. Using~\eqref{eq:trs},
the cubic term remains invariant:
\begin{equation}
\text{Tr}\, \bar{M}^3 \to \text{Tr}\, 
(\Lambda^T \bar{M}^{T} \Lambda)^3 = \text{Tr} (\bar{M}^T)^3 = \text{Tr} \,\bar{M}^3. 
\end{equation}
Therefore, it cannot be ruled out by TRS.  To see that it vanishes only in the SU($2$) case, we must use  that $\bar{M} = \sum_{a=1}^{3} m_a \sigma^a/2$,
$\Lambda = -i \sigma^2$, therefore $\Lambda^{T} \bar{M}^{T
} \Lambda  = -\bar{M}$ because of the anti-commutation properties of the Pauli matrices and $(\sigma^2)^T = -\sigma^2$. Thus, $\text{Tr}\, M^n \to -\text{Tr} M^n$ under TRS for $n$ odd. Note that $\Lambda\propto \sigma^2$ (i.e. one of the generators of the Lie algebra)is a property specific to SU($2$) which is not shared by the higher rank SU($N>2$) groups. In the latter case, the traceless matrix $\Lambda$ is a linear combination of the generators of the Lie algebra $T^a$ and therefore $\Lambda^{T} \bar{M}^{T} \Lambda \neq -\bar{M}$.

\section{Unrestricted Hartree-Fock}\label{sec:uhf}
\begin{figure}[t]
\center
\includegraphics[width=\columnwidth]{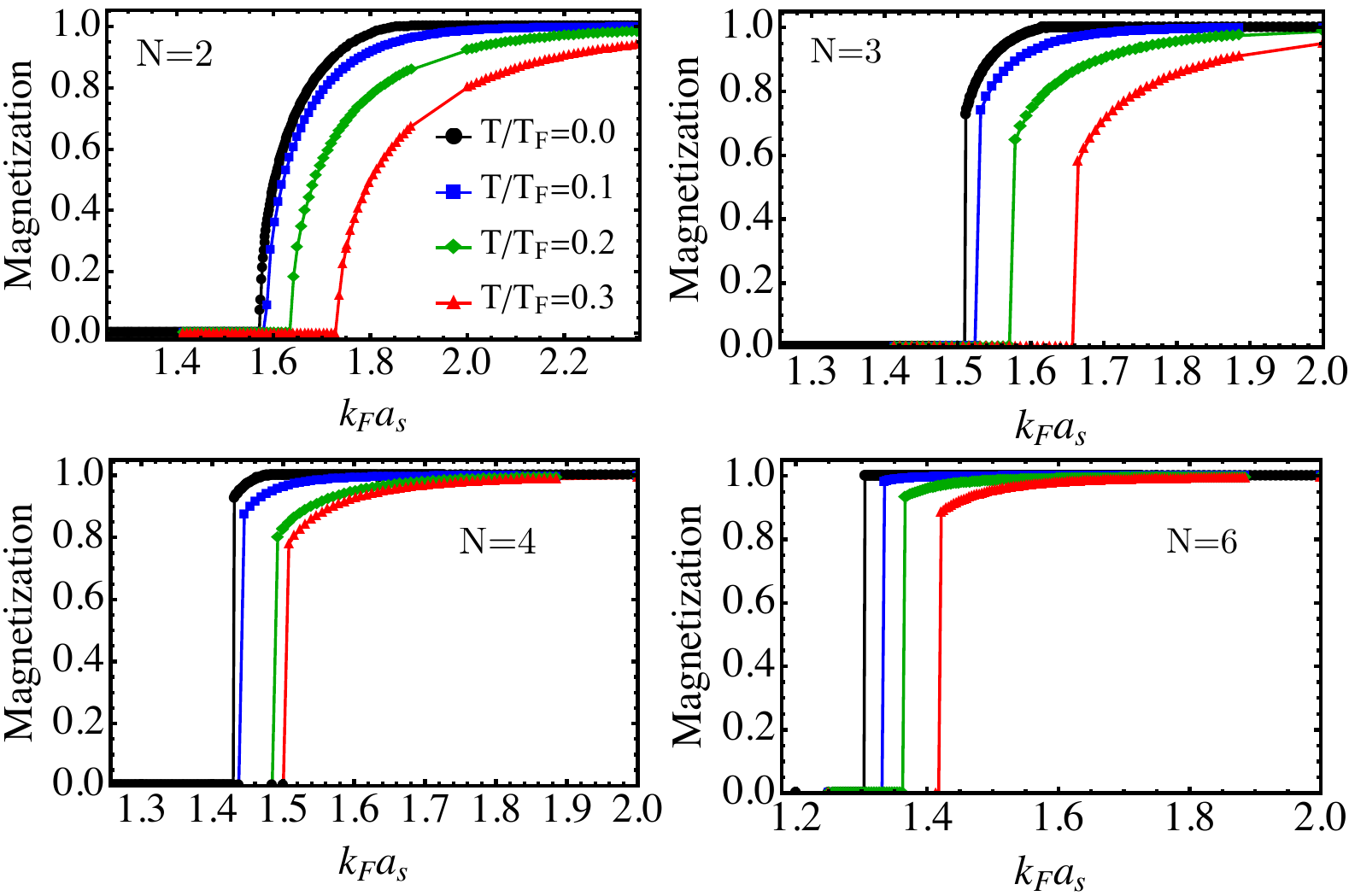}
\caption{Scalar magnetization (see definition in Eq.~\eqref{eq:mag}) as a function of the dimensionless gas parameter $k_F a_s$ ($k_F$ is the Fermi momentum and $a_s$ the s-wave scattering length). Results obtained using the unrestricted Hartree-Fock method. Note the finite jump that is present at all the studied temperatures ($T\leq 0.3\: T_F$, $T_F$ being the Fermi temperature). In the latter case, the magnetization rises continuously for $k_F a_s \leq \pi/2$
. Within mean-field theory, the transition becomes more abrupt with increasing $N> 2$ and the scalar magnetization rises discontinuously for values of the gas parameter smaller than the value dictated by Stoner's criterion, i.e. $k_F a_s = \pi/2$.} \label{HF}
\end{figure}

In order to make contact with an experimentally relevant system, below we consider a three-dimensional Fermi gas with SU($N$) symmetry, which is described by the following Hamiltonian~\cite{Cazalilla_NJOP_11_2009} (in units where $\hbar = 1$):
\begin{align}
H &=H_0 + H_{\text{int}},\notag\\
H_0 &= \int d\vec{r} \, c^{\dag}_{\alpha}(\vec{r})\left(-\frac{\vec{\nabla}^2}{2m}-\mu_{\alpha}\right)c^{\alpha}(\vec{r}), \notag\\
H_{\text{int}}&= \frac{g}{2\Omega} \int d\vec{r} \, c^{\dag}_{\alpha}(\vec{r})c^{\dag}_{\beta}(\vec{r})c^{\beta}(\vec{r}) c^{\alpha}(\vec{r}).\label{eq:H}
\end{align}
In this expression and those to follow (unless stated otherwise) we  use Einstein's repeated (Greek) index summation convention.  The above Hamiltonian is written in terms of a set of Fermi field operators $c^{\alpha}(\vec{r})$ which transform according to the fundamental representation(s) of SU($N$) and obey
\begin{equation}
\left\{c^{\alpha}(\vec{r}), c^{\dag}_{\beta}(\vec{r}^{\prime}) \right\}=\delta(\vec{r}-\vec{r}^{\prime}) \: \delta^{\alpha}_{\beta},
\end{equation}
anti-commuting otherwise. The gas is confined in a volume $\Omega$  and we assume periodic boundary conditions. The s-wave contact interactions are described by a  pseudo-potential interaction parameter $g \propto a_s$ (to lowest order, $a_s$ being s-wave scattering length). The Hamiltonian exhibits SU$(N)$ symmetry when all $\mu_{\sigma}$ are equal. Different $\mu_{\alpha}$
are necessary in order to describe  magnetized states.
This model describes an ultracold gasof Akaline-Earth atoms which exhibits an emergent SU$(N)$ symmetry~\cite{Cazalilla_NJOP_11_2009,Gorshkov_2010,Cazalilla_2014}. 

 We have performed unrestricted Hartree-Fock (UHF) calculations for the above model~\eqref{eq:H} by minimizing at constant particle density and temperature the expression of the grand canonical free-energy computed to first order in the interaction. As described in the Appendix,  the grand canonical free-energy in the Hartree-Fock approximation reads:
 \begin{equation}
 F_{HF} = F_0 + F_1,
 \end{equation}
where $F_0= -\frac{1}{\beta_T}\sum_{\vec{k},\alpha}\log[1+e^{-\beta_T(\epsilon_{\vec{k}\alpha}-\mu_\sigma)}]$ is the free-particle free energy and  
\begin{align}\label{eq:F1}
F_1= \left(\frac{2\pi a_s}{m} \right) \sum_{\alpha\neq\beta}
\sum_{\vec{q k}} n_{\vec{q}\alpha} n_{\vec{k}\beta},
\end{align}
the Hartree-Fock energy. The Fermi-Dirac distribution functions (no summation over repeated indices implied here)  are $n_{\vec{k},\alpha} = \langle c^{\dag}_{\vec{k}\alpha} c^{\alpha}_{\vec{k}}\rangle = \left( e^{\beta_T (\epsilon_{\vec{k}}-\mu_{\alpha})} +1 \right)^{-1}$ for $\alpha = 1, \ldots, N$ at  absolute inverse temperature $\beta^{-1}_T = k_B T$ and the chemical potential for each component $\mu_{\alpha}$. The latter must be found self-consistently while keeping  the total particle  density,
\begin{equation}
n_{\text{tot}} = \frac{N_{\text{tot}}}{\Omega} = \sum_{\vec{k},
\alpha} n_{\vec{k},\alpha},
\end{equation}
constant.
Unlike Landau's mean-field theory, the UHF method does not rely on the assumption that the order parameter is small in the neighborhood of the phase transition.  However, we shall see that this assumption of Landau's  theory is not essential and the latter correctly captures the order of the transition.   

The results of minimizing the energy following the UHF method  are shown in Fig.~\ref{HF} for  $N=2, 3, 4$, and $N=6$.  We numerically find that SU($N$) symmetry is broken in a pattern where one of the species ``cannibalizes'' the $N-1$ others.  Thus the value of the order parameter in broken symmetry state is fully characterized by a single scalar "magnetization'' parameter. Since, by adiabatic continuity, Landau quasi-particles carry the same charge and SU($N$) quantum numbers as the original constituent fermions the order parameter can be obtained from the single-particle density matrices as follows:
\begin{equation}
\bar{M}_{\beta}^{\alpha} = 
\int d\vec{r}\, \left[ \langle c^{\dag}_{\beta}(\vec{r})c^{\alpha}(\vec{r}) \rangle - \frac{\delta^{\alpha}_{\beta}}{N}  \langle
c^{\dag}_{\gamma}(\vec{r})c^{\gamma}(\vec{r}) \rangle \right].
\end{equation}
Choosing the (majority) component that grows at the expense of the other $N-1$ (minority) components to be $\alpha=N$,  the scalar polarization  of the system is defined as follows:
\begin{align}
m_s &=  \frac{1}{N_{\text{tot}} }  \int d\vec{r}\: \left[   
  \langle c^{\dag}_{N}(\vec{r})
c^{N}(\vec{r})\rangle  -  \langle c^{\dag}_{N-1}(\vec{r})
c^{N-1}(\vec{r})\rangle 
\right] \notag \\
&=\frac{1}{N_{\text{tot}} }  \int d\vec{r}\: \left[    \langle c^{\dag}_{N}(\vec{r})
c^{N}(\vec{r})\rangle  -  \frac{1}{N-1}\sum_{\alpha=1}^{N-1} \langle c^{\dag}_{\alpha}(\vec{r})
c^{\alpha}(\vec{r})\rangle 
\right] \notag \\
&= \frac{1}{N_{\text{tot}}}\sqrt{\frac{2N}{N-1}} \text{Tr} \left[  T_{N^2-1} \bar{M} \right], \label{eq:mag}
\end{align}
where
\begin{align}
T_{N^2-1} = \frac{1}{\sqrt{2N(N-1)}} \left(
\begin{array}{ccccc}
-1 & 0 & \cdots & 0 &0\\
0 & -1 & \cdots & 0 & 0 \\
\vdots & \vdots  & \cdots & \vdots & \vdots\\
0  & 0 & 0 & -1 & 0\\
0 & 0 & 0 & 0  & N-1
\end{array}
\right)
\end{align}
is the diagonal Lie algebra generator with the largest number of non-zero matrix elements.

  In Fig.~\ref{HF},  $m_s$  is plotted as a function of the dimensionless gas parameter $k_F a_s$  ($k_F$ being the Fermi momentum and $a_s$ the s-wave scattering length characterizing the interaction) for different values of the temperature from $T=0$ up to $T=0.3\: T_F$, ($T_F  = E_F/k_B$ is the Fermi temperature, $E_F = k^2_F/2m$ being the Fermi energy). 

 On the other hand, for  the SU($N=2$)-symmetric system, the magnetization grows continuously from zero at all the calculated temperatures, also in agreement with Landau's theory as described above. It is worth  emphasizing  that the predictions of Landau's theory remain qualitatively correct despite the  large  jump in the magnetization that is numerically observed. 
Indeed, the transition appears to become more abrupt as $N$ increases: Compared to $N=3$, the system with $N=6$ reaches the fully polarized state a smaller values of  $k_F a_s$. Moreover, 
even for temperatures as high $T=0.3\: T_F$, the gas parameter regime where a partially polarized system exists is rather small. 

\section{Fluctuation Effects}\label{sec:2nd_p}

As mentioned above,  particle-hole excitations couple
to the magnetization and, by a mechanism akin to the Weinberg-Coleman mechanism of particle
Physics~\cite{WeinbergColeman_PhysRevD.7.1888}, the continuous phase transition predicted by Landau's theory becomes a first order  transition. The phenomenon has been discussed in depth
in the literature (e.g. Refs.~\cite{Belitz_PhysRevLett.89.247202_2002,Belitz_Kirkpatrick_PhysRevLett.94.247205,
Duine_PhysRevLett.95.230403_2005,Conduit_PhysRevA.79.053606_2009,Conduit_PhysRevA.83.043618_2011} and references therein) in connection with the breakdown Hertz-Millis-Moriya theory of 
quantum criticality in  metallic ferromagnets~
\cite{Hertz_PhysRevB.14.1165}.

When trying to capture the effects of particle-hole fluctuations, several authors have pointed out for the two-component Fermi gas~\cite{Duine_PhysRevLett.95.230403_2005,Conduit_PhysRevA.79.053606_2009,Conduit_PhysRevA.83.043618_2011} that minimization of the expression of the free-energy  to second order in $k_F a_s$ is sufficient. Below, we have generalized this approach to SU($N > 2$). Although it  may not be sufficiently accurate in order to determine  the transition point ($k_F a_s \approx 1$ is likely beyond the validity  of the  second-order free-energy), 
the method allows to get a rough estimate of the corrections to the mean-field 
theory  arising from particle-hole fluctuations. 
Indeed, we find that, contrary to what a (second-order) perturbative calculation of the magnetic susceptibility of the paramagnetic gas suggests~\cite{Yip_PhysRevA.89.043610}, second order corrections to the UHF method enhance the ferromagnetic tendencies and make the first order transition  even more abrupt. 

\begin{figure}[t]
\center
\includegraphics[width=\columnwidth]{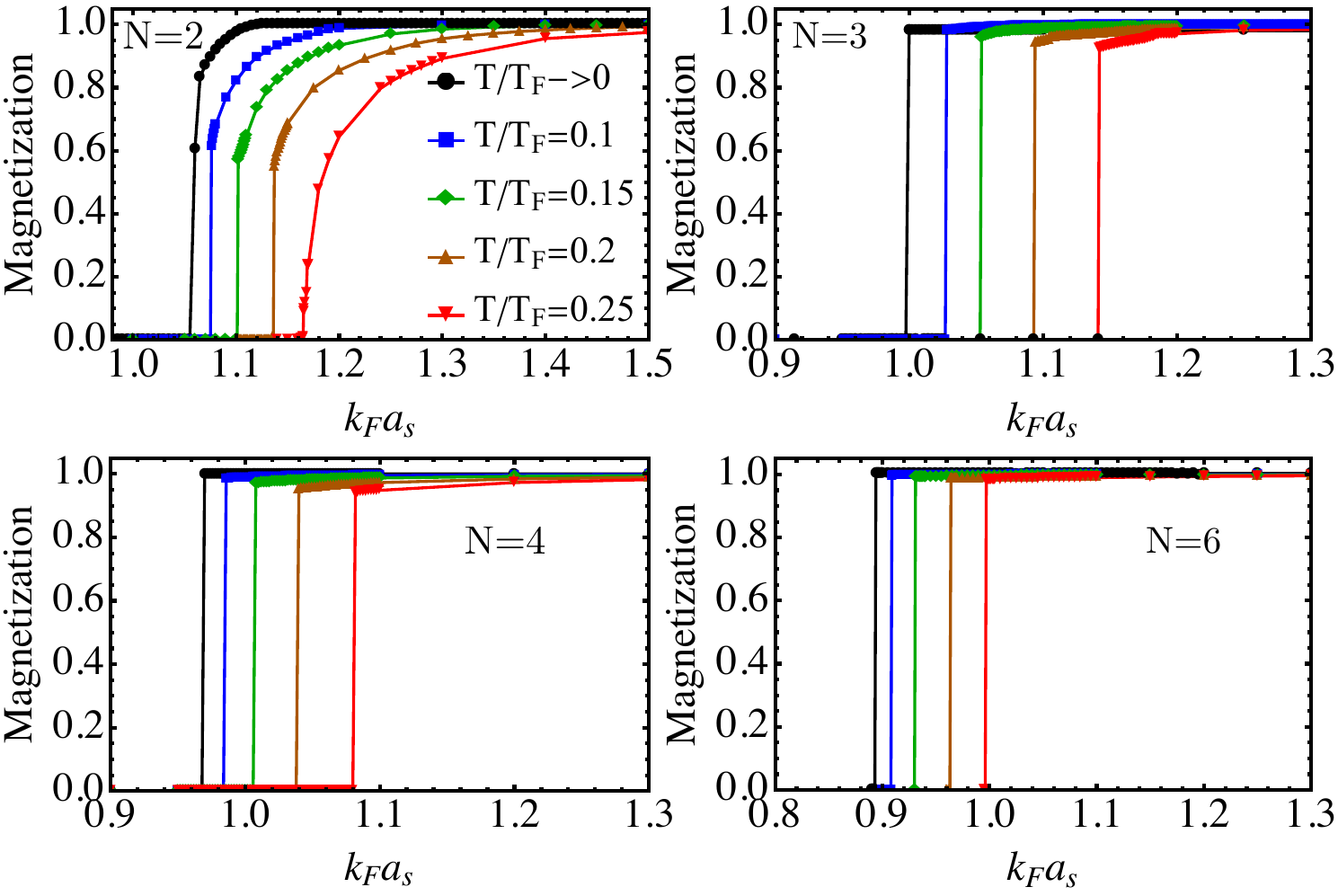}
\caption{Scalar magnetization (see Eq.~\eqref{eq:mag} for definition) obtained by minimizing the second-order expression of the free energy  as a function of the gas parameter $k_F a_s$ for $N=2, 3, 4, 6$ and for temperature $T$ ranging from $T = 0$  to $T=0.25 T_F$ ($T_F$ is the Fermi temperature).
 For $N = 2$, the ferromagnetic transition becomes  first order for $T\leq T_{tcp}$, where  $T_{\text{tcp}} \approx 0.2 \: T_F$  is the tri-critical temperionsature. Above  $T_{\text{tcp}}$, thermal flutuations smear the Fermi surface and the transition becomes second order. On the other hand, for $N>2$  the Fermi  surface fluctuations captured by the second order correction to the free energy make the transition even more abrupt at all temperatures.}
 \label{snd}
\end{figure}
\begin{figure}[t]
\center
\includegraphics[width=\columnwidth]{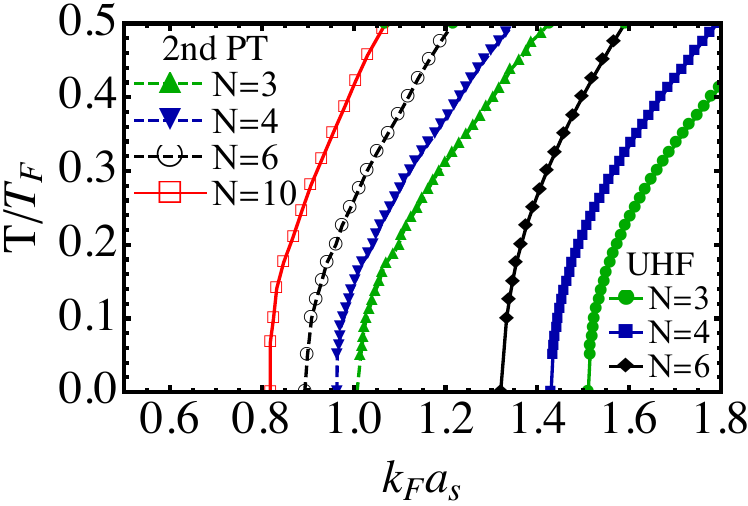}
\includegraphics[width=\columnwidth]{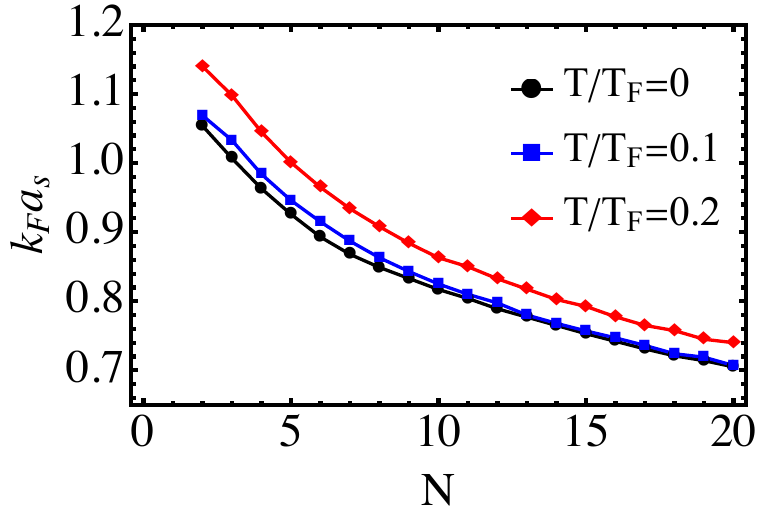}
\caption{ (top) Numerically determined phase boundary of the (first order) phase transition between the paramagnetic gas and the itinerant ferromagnet at different values of the ratio of the temperature to the Fermi temperature ($T/T_F$) for $N>2$. The left four curves are results of second order perturbation theory. The right three curves are derived from mean-field theory. (bottom) Value of the transition gas parameter obtained using  second order perturbation theory for different temperatures as a function of the number of components $N$ of the SU($N$)-symmetric Fermi gas. }
 \label{bd}
\end{figure}

  Using the method described above requires that we minimize the two leading corrections to the free energy. The details of their calculation are found in Appendix~\ref{app:pert}. The expression of the free energy subject to unconstrained minimization is $F= F_0+F_1+F_2$, where
\begin{align}\label{eq:F2}
F_2= -\frac{1}{2} \left(\frac{4\pi a_s}{m\Omega}\right)^2  \sum_{ \alpha\neq\beta} \sum_{ \vec{pkq} } 
\frac{ n_{\vec{p}\alpha} n_{\vec{k}\beta}
(n_{\vec{p+q} \alpha}+n_{\vec{k-q} \beta}) }{\epsilon_\vec{p}+\epsilon_\vec{k}-\epsilon_{\vec{p+q}}-\epsilon_{\vec{k-q}}}
\end{align}
is the second order correction after renormalization~\cite{Abrikosov1965,Huang_PhysRevA.99.063612_2019,Huang_Yang_PhysRev.105.767,Parthia2011} (see also Appendix ~\ref{app:pert} for details).  

The results for the scalar magnetization of the minimization of the second order free-energy as a function of the gas parameter $k_F a_s$ are shown in Fig.~\ref{snd}. As mentioned above,  fluctuation effects induce a first order first transition at $T \leq T_{
\text{tcp}}$ ($T_{\text{tcp}}\approx 0.25\:  T_F$)~\cite{Duine_PhysRevLett.95.230403_2005,Conduit_PhysRevA.79.053606_2009}  for SU($N= 2$). By contrast,  particle-hole fluctuations  only makethe first order transition even more abrupt in SU($N>2$) symmetric systems at all studied temperatures.  

In Fig.~\ref{bd} (top panel), we show the first-order phase boundary between the paramagetic gas and the itinerant ferromagnet for SU$(N = 3, 4, 6)$, up to $T = 0.5\: T_F$. Both the boundary obtained using the UHF and  minimization of the  free-energy up to second order are shown. Notice that the overall effect of  fluctuations is to shift the phase boundary to lower  values of  $k_F a_S$. In the studied temperature range,  we did not find a tri-critical point like the one observedfor SU($N=2$). This confirms the expectation that the first order character of the transition is already well captured by  mean-field theory and fluctuations only enhance it.  

 In the lower panel of Fig.~\ref{bd} we show 
the phase boundary as a function of the number of components $N$ and  temperature, as obtained by minimization of the free energy up to second order in $k_F a_s$. Note the general trend that larger $N$ shifts the boundary to lower values of $k_F a_s$, although caution must be exercised as the boundary lies in a region where $k_F a_s \sim 1$, which is beyond the validity of the second-order free-energy. Nonetheless, it is worth pointing out that the overall trend of shifting the transition to lower $k_F a_s$ is also  captured by the UHF calculations.

Finally, let us point out that we have checked that numerical integration of the second order free-energy  yields the same results at $T=0$ as integration using the formula obtained by
Kanno~\cite{Kanno_PTP_1970}. The reader is referred to Appendix~\ref{app:kanno} for details. 

\section{Finite temperature Tan's contact}\label{sec:tan}

\begin{figure}[t]
\center
\includegraphics[width=\columnwidth]{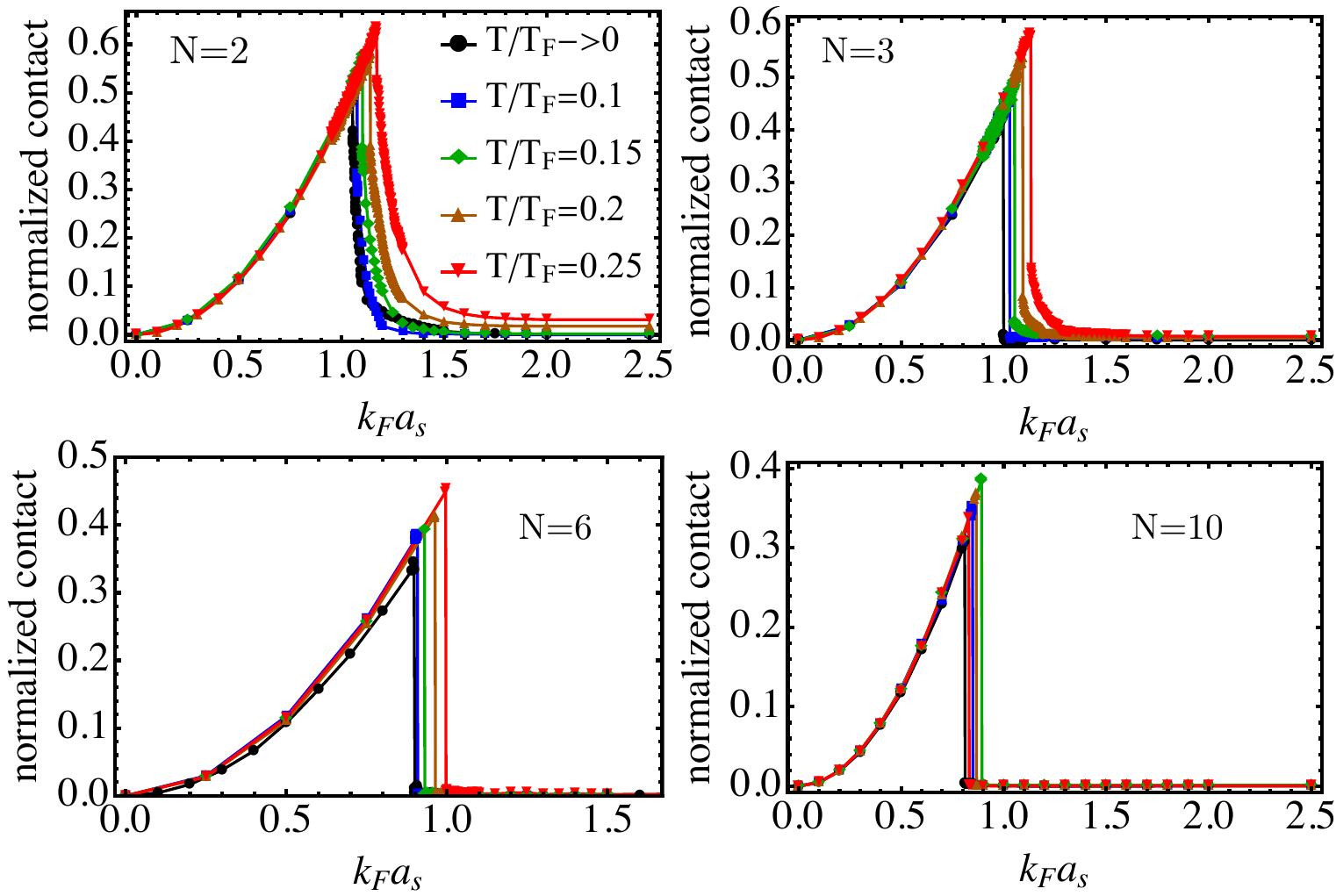}
\caption{Leading order contribution ($O[(k_Fa_s)^2]$) to Tan's contact scaled by $(n_{\text{tot}}/N)^{2-2/d} [N(N-1)]$ for an SU($N$)-symmetric Fermi gas at different temperatures. In the paramagnetic phase, all the curves collapse to $\propto(k_Fa_s)^2 (n_{\text{tot}}/N)^{2-2/d}N (N-1)$ with $d=3$ is the dimensionality of system.   In ferromagnetic phase, thermal fluctuations make the  the contact finite even in the FM phase but this effect is suppressed with increasing $N$.}\label{fig:cts}
\end{figure}
 An experimentally accessible observable that has attracted much attention in recent times in Tan's contact~\cite{TAN_AnnPhys_2008,Huang_PhysRevA.99.063612_2019}. This quantity is defined as the coefficient  of the $k^{-4}$ term the momentum distribution at large momentum $k\gg k_F$.
 
  Using the same methods employed to compute second order free-energy (see e.g.~\cite{Huang_PhysRevA.99.063612_2019}), we can obtain the momentum distribution and extract Tan's contact from the resulting expression.  The leading order correction is second order in the gas parameter $k_F a_s$ and for species $\alpha$ reads:
\begin{multline}
\delta N_{\vec{k}\alpha}=\frac{2}{\Omega^2}\left(\frac{4\pi a_s}{m}\right)^2\sum_{\vec{qr},\beta\neq \alpha}\left[n_{q\alpha}n_{r\beta}-n_{\vec{k}\alpha}n_{|\vec{q}+\vec{r}-\vec{k}|,\beta}\right]\\
\times
\left[\frac{k^2}{m}\left( 1+\frac{\vec{q}\cdot\vec{r}}{k^2}-\frac{\vec{k}\cdot(\vec{q}+\vec{r})}{k^2}\right)\right]^{-2}+O[(k_Fa_s)^3],
\label{eq:Cneq}.
\end{multline}
Hence, Tan's contact  at finite temperature is obtained from the limit  $C =\lim_{k \to +\infty} k^4\sum_\sigma \delta N_{\vec{k}\sigma}$.
Using the above expression, we can
write the result as follows:
 \begin{align}
C(T) &=\frac{4}{\Omega^2}\left(\frac{4\pi a_s}{m}\right)^2\sum_{\vec{qr},\alpha\neq\sigma}n_{\vec{q}\sigma}n_{\vec{r}\alpha}+\delta C(T)+O(a_s^3)
 \label{eq:Ceq}
\end{align}
Here $\delta C(T)$ is the contribution of the second term in Eq.~\ref{eq:Cneq}, which yields an exponentially small contribution to the contact. 

Turning our attention to the first term in Eq.~\ref{eq:Ceq}, we notice that it is proportional to
\begin{equation}
C(T) \propto \left(\frac{4\pi a_s}{m}\right)^2 \sum_{\alpha\neq\beta} n_{\alpha}(T) n_{\beta}(T) + O(a^3_s)
\end{equation}
where $n_{\alpha} = \sum_{\vec{k}} n_{\vec{k},\alpha}(T)$ is the density of  component $\alpha = 1, \ldots, N$. We can rewrite the sum in the above equation in terms of the expectation value of the square of the diagonal generators of SU($N$):
\begin{equation}
\sum_{\alpha\neq\beta} n_{\alpha}(T) n_{\beta}(T) = \frac{N-1}{2N} n^2_{\text{tot}} - \frac{1}{\Omega^2}
\sum_{r=2}^{N} \left( \langle T_{r^2-1} \rangle \right)^2
\end{equation}
Note that the first term $\propto n^2_{\text{tot}}$
is independent of temperature as the number of
particles is fixed. Therefore, all the temperature dependence stems from the second term, i.e. the sum over the diagonal generators of SU($N$). For the system of interest here, the  Fermi gas with short-range repulsive interactions, as described above, the only generator whose expectation value changes across the transition is $T_{N^2-1}$, which implies that
\begin{align}
C(T) - C(T=0) &\simeq - \left(\frac{4\pi a_s}{m \Omega}  \langle T_{N^2-1} \rangle \right)^2  \\
&\propto - \left(\frac{4\pi a_s}{m}
\right)^2 m_s^2(T),
\end{align}
where $m_s(T)$ is the scalar magnetization at temperature $T$. The above result allows us to understand the behavior of the contact around the transition to the itinerant ferromagnet.
Thus,  in the paramagnetic gas phase (itinerant ferromagnet) where $m_s(T) = 0$ ($m_s(T) = 1$), the (leading order) contact does not depend on $T$ (up to exponentially small corrections arising from $\delta C(T)$). However, in the neighborhood of the transition, its temperature dependence is controlled by   $m_s(T)$. For $N=2$, the existence of the tri-critical point leads to a stronger temperature dependence than for $N > 2$. In the latter case, the abrupt  first order transition described above  strongly reduces the parameter regime where a partially polarized phase exists, which translates in a sharper decrease of Tan's contact as the system transitions to the itinerant ferromagnet.

Besides the temperature dependence, as shown in Fig.~\ref{fig:cts},  in the paramagnetic gas phase, Tan's contact scales as $C\propto c_d(k_Fa_s)^2 (n_{\text{tot}}/N)^{2-2/d}N(N-1)$ where $d = 3$ is dimensionality and $c_d$ is a constant. Therefore, it increases as $(k_F a_s)^2$ as a function of the gas parameter. This is clearly an artifact of the perturbative approach used to obtain Eq.~\eqref{eq:Cneq}.

 In general, the contact will exhibit a more complicated behavior with $k_F a_s$ and $T$ than described above. Nevertheless, despite the transition happening for $k_F a_s \approx 1$, which is beyond the applicability of 
 second-order perturbation theory, we expect the conclusions of the above analysis to remain qualitatively correct, and  Tan's contact for an SU($N = 2$)-symmetric gas to exhibit a stronger temperature dependence around the transition  due to the presence of the tri-critical point.  Finally, it is worth mentioning that there has been already a preliminary exploration of the temperature dependence of Tan's contact in Ref.~\cite{BO_PhysRevX.10.041053_2020}. However, in this work the gas  is  in a high temperature regime where the temperature dependence of Tan's contact is fairly  well reproduced by the virial expansion (accounting for trap effects). This high-temperature regime is different from the Fermi liquid regime investigated here.

\section{Discussion and Conclusions}\label{sec:conclusion}

 The results of the unrestricted Hartree-Fock (UHF) calculations discussed in previous sections
 confirm  the picture provided by Landau's  theory  of phase transitions of a first order transition to the itinerant ferromagnet.
 Beyond mean-field theory, fluctuation effects as estimated by minimizing  the  free-energy up to second order in $k_F a_s$ make the first order  transition more abrupt. Generally speaking, for $N > 2$ fluctuations lead to a fully polarized Fermi gas at lower values of the gas parameter $k_F a_s$ and make the fully polarized gas stable at higher temperatures. For the SU($N=2)$ Fermi gas, we have also reproduced results obtained earlier in Refs.~\cite{Duine_PhysRevLett.95.230403_2005,Conduit_PhysRevA.79.053606_2009}. In particular, we have observed the existence of a tri-crical point~\cite{Belitz_PhysRevLett.89.247202_2002,Duine_PhysRevLett.95.230403_2005,Conduit_PhysRevA.79.053606_2009} at a temperature $T_{tcp}\simeq 0.2 \: T_F$. Above $T_{tcp}$ thermal fluctuations smooth out the Fermi surface and the transition to the itinerant FM state becomes continuous. Above we have  pointed out the that the presence of the tri-critical point leads to a stronger dependence on temperature of Tan's contact for $T/T_F \lesssim 0.25$ in $N=2$ component Fermi gases.

 In Ref.~\cite{Cazalilla_NJOP_11_2009},
Stoner's criterion was found to be $k_F a_s = \pi/2$, independent of the number of components, $N$. However, both the UHF and the minimization of the second order free-energy
show that the transition point depends on $N$.
We may be tempted to think that this is because the quoted result was derived using the lowest order (HF) approximation to the Landau Fermi liquid parameter $F^m_0$~\cite{Cazalilla_NJOP_11_2009}. Indeed, higher order corrections computed within Fermi liquid theory in the paramagnetic state~\cite{Yip_PhysRevA.89.043610} do not  improve the accuracy of the estimate but make things worse: As shown in  Ref.~\cite{Yip_PhysRevA.89.043610} to $O[(k_F a_s)^2]$, the inverse magnetic susceptibility for an SU($N$)-symmetric Fermi liquid with contact interactions reads:
\begin{align}
\frac{\chi^{-1}_m}{\left(\chi^0_m
\right)^{-1}}  &= 1 - \frac{2k_F a_s}{\pi
}  - \frac{8 (k_F a_s)^2}{15\pi^2} \left[ \left(11-\frac{7 N}{2} \right) \right. \notag\\
&\quad \left. + 2(N-1)\log 2 \right] + O[(k_F a_s)^3]
\end{align}
where $\chi^0_m$ is the free Fermi gas susceptibility.  In this approximation, $\chi^{-1}_m$ changes sign at values $k_F a_s >\pi/2$ for $N < 6$ and has no change of sign for $N \geq 6$, thus predicting no itinerant FM transition at all.  Hence, we may conclude that the various low order approximations to the zero field  susceptibility  are not very informative about the FM tendencies of SU$(N>2$) Fermi liquids. However, since we know that we are dealing with a first order phase transition that requires a large fluctuation for the system to reach the global minimum of the (free) energy,  a more promising  approach should be a comparison of the total energies of the fully polarized and paramagnetic Fermi gases.
Indeed, for $N = 2$ this was already was noticed by Morita \emph{et al.}~\cite{Morita_10.1143/PTP.18.326} as early as 1957: These authors noticed that the second order correction to the ground state energy of the paramagnetic gas obtained by Huang-Yang~\cite{Huang_Yang_PhysRev.105.767} is positive. On the other hand,  the energy of the fully polarized state is independent of $k_F a_s$, which should lead to a FM ground state at sufficiently large $k_F a_s$. In a field theoretic language, the positive sign of the second order correction is a consequence of the renormalization of the interaction coupling (see Appendix~\ref{app:pert}). Indeed, the positivity of the second order correction to ground state energy  holds for all $N \geq 2$. 

\begin{figure}[t]
\center
\includegraphics[width=\columnwidth]{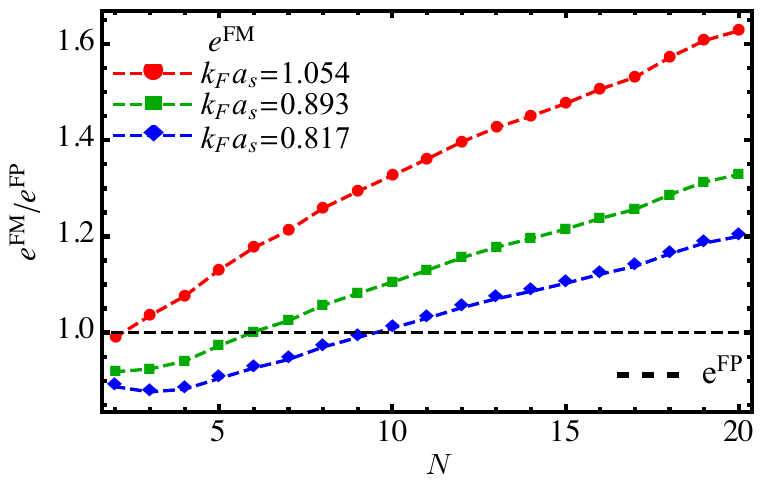}
\caption{Comparison of the ground state energy of the fully polarized Fermi gas (black dashed horizontal line) and paramagnetic Fermi gas (color dashed lines with markers) for different values of the gas parameter $k_F a_s$. Wherever the curves for the energy of the paramagnetic gas cross the 
horizontal line of the fully polarized gas  approximately corresponds to the value of $k_F a_s$ of the first order transition. Note that by decreasing the interaction from $k_F a_s \simeq 1$, the lines cross at values of $N > 2$.}
 \label{sc}
\end{figure}

 Thus, we can compare the ground state energies in order to understand how the transition point shifts with $N$. To this end,  let us  consider the scaling  with $N$ of the different contributions to the ground state energy at constant $k_F a_s$. We find: 
\begin{align}
e^{\text{FM}}_0 &= c^d_0 N^{-1-2/d},\\
e^{\text{FM}}_1 &=  c^d_1 (k_Fa_s) N(N-1) N^{1/d-2}_c, \\
e^{\text{FM}}_2&=  c^d_2 (k_Fa_s)^2 N(N-1)N^{2/d} f_2(N),
\end{align}
where $d = 3$ is the space dimensionality and  $c^d_{i=1,2}$ are  functions of $d$ and 
$k_F$ and $f_2(N)$ is a function of $N$ that needs to be evaluated numerically. However, for fully-polarized phase, the total energy equals the kinetic energy and 
it is independent of $N$ and in Fig.~\ref{sc} it appears as a horizontal line. On the other hand,  for the paramagnetic gas, the total energy  monotonically increases with $N$ (cf.  Fig.~\ref{sc}). The crossing of these curves with the horizontal line agrees well with the value of $N$ at which the transition to the itinerant FM takes place at $T=0$. Thus we see the critical gas parameter decreases with increasing $N$, in agreement with the unconstrained minimization results reported in previous sections.  

 Let us also comment on the differences  in the effects of fluctuations between  SU($N=2$) and SU($N >2$). The existence of a cubic term in the Landau free energy has important consequences when considering the effect of fluctuations. As explained in Ref.~\cite{Belitz_PhysRevLett.89.247202_2002} it is possible to capture the effect of fluctuations by using a generalized Landau free energy, which (at zero temperature) contains a non-analytic term $m^4_s \log \: m^2_s$, where $m_s$ is the scalar magnetization. Qualitatively such non-analytic term  can modify the sign of the  (positive) quartic term $\sim m^4_s$ in the Landau free energy. This modification is most important when the system approaches criticality and  $m_s$ mainly fluctuates around zero thus leading to a substantial negative correction to the quartic term. On the other hand, for SU($N > 2$), the existence of the cubic term means  transition to the itinerant ferromagnet is driven by large fluctuations and the effect of the $m^4_s \log \: m^2_s$ is less important compared to $N = 2$. 

 As above,  the results reported here apply to a uniform interacting Fermi gas in three spatial dimensions. Let us briefly consider the situation in two spatial dimensions, which turns out to be somewhat different in several respects. Application of the results obtained in  Ref.~\cite{Cazalilla_NJOP_11_2009} for the Landau free energy  show that the coefficients of all the terms proportional to 
 $\mathrm{Tr}\: \bar{M}^n$ with $n > 2$ vanish because they are proportional to  derivatives of the (non-interacting) density of states, which is constant two spatial dimensions (for quadratic single-particle dispersion). Thus, the Landau free energy simply contains the term $\propto \mathrm{Tr}\: \bar{M}^2$ and when the coefficient of the second order term ($c_2$ in Eq.~\ref
{eq:landau}) changes sign for sufficiently strong interaction, the free energy is minimized by 
 letting the system become fully polarized. Applying the unrestricted Hartree-Fock method yields the same  pattern of SU($N$) symmetry breaking as for the three-dimensional gas (i.e. SU$(N) \to$ SU($N-1$)$\times$ U($1$)). When coupling to particle-hole fluctuations is accounted for, we expect a situation for SU($N>2$) similar to the one reported in Ref.~\cite{Conduit_PhysRevA.82.043604} for SU($N=2$) gas, namely the non-analytical corrections to the free energy to lead to tri-critical point where the order transition changes from first to second. Further details of the FM transition of the uniform gas in two spatial dimensions will be reported elsewhere~\cite{unpub}. 

Finally, a few words about experiments are in order. Currently,  it is difficult that experiments using alkaline-earth ultracold gases with emergent SU($N> 2$) symmetry can  reach the regime where the gas parameter  $k_F a_s\sim 1$. Even with the decrease in the critical value of the gas parameter achieved with large values of $N$, the regime of where the transition to the itinerant ferromagnet
can be reached seems hard to approach as standard (magnetic-field tuned) Feshbach resonances are not available for alkaline-earth atoms in their ground state~\cite{Cazalilla_2014}. On the other hand,
using an optical Feshbach resonance~\cite{Enomoto_PhysRevLett.101.203201} to enhance the interaction means the SU$(N)$ symmetry will be broken by the interaction. Thus, our theoretical description should account for deviations for SU($N$) symmetry and, more importantly, for radiative losses and the presence of lower-energy molecular channels. Dealing with such extensions to the theory reported here is a very interesting and challenging problem which will be tackled in the future as experiments hopefully begin to explore the strongly interacting regime. 

\acknowledgments

 This work has been supported by Ikerbasque, Basque Foundation for Science (MAC) and MICINN Grant No. PID2020-120614GB-I00 (ENACT).
 CHH acknowledges a PhD fellowship from the Donostia International Physics Center (DIPC).   We thank Chunli Huang for useful discussions. 
 
\appendix

\section{Transformation of the order parameter under time-reversal}

 Let us denote the operator for time reversal transformations as $\mathcal{T}$ and recall the transformation rule of a spin-$F$ 
 Fermi field.  For a spin-$F$ fermion ($F$ being half-integer), the operator $\mathcal{T} = e^{-i \pi \hat{F}^y} \mathcal{K}$ where  $\mathcal{K}$ is an anti-unitary operator which turns $i\to -i$, 
 and $\hat{F}^y$ is the $y$-component of the spin operator. 
Thus, arranging the components of the spin in decreasing order of  $F^z$ in a $2F+1$-component spinor $\Psi(\vec{r})$, we have
\begin{equation}
   \mathcal{T}\Psi^{\alpha} \mathcal{T}^{-1} = \Lambda^{\alpha}_{\beta} \Psi^{\beta} 
\end{equation}
where $\Lambda = e^{i\pi F^y}$ is the matrix displayed in Eq.~\eqref{eq:lambda}.
Note that in the case of $F=1/2$, $\Lambda = -i \sigma^y$ (where $\sigma^y$ is a Pauli matrix) and
therefore  we recover the well known result: $\mathcal{T} c^{\uparrow}(\vec{r}) \mathcal{T}^{\dag} = -c^{\downarrow}(\vec{r})$ and $\mathcal{T} c^{\downarrow}(\vec{r}) \mathcal{T}^{\dag} = c^{\uparrow}(\vec{r})$. 

 The (nuclear=total) spin operator in second quantized form 
\begin{equation}
\vec{\hat{F}} =  \int d\vec{r}\,
c^{\dag}_{\alpha}(\vec{r})  \vec{F}^{\alpha}_{\beta} c^{\beta}(\vec{r}).
\end{equation}
transforms according to
\begin{align}
\mathcal{T} \vec{\hat{F}}  \mathcal{T}^{-1} &=    \int d\vec{r}\,  (F^{\alpha}_{\beta})^{*}
\mathcal{T} c^{\dag}_{\alpha}(\vec{r}) \mathcal{T}^{-1} \mathcal{T}  c^{\beta}(\vec{r}) \mathcal{T}^{-1}  \\
&=   \int d\vec{r}\,  (F^{\alpha}_{\beta})^{*}
\Lambda^{\gamma}_{\alpha} c^{\dag}_{\gamma}(\vec{r})  \Lambda^{\beta}_{\delta} c^{\delta}(\vec{r}) \\
&=\int d\vec{r} \, c^{\dag}_{\gamma}(\vec{r})
) \left( \Lambda^{T} F^T  \Lambda \right) c^{\delta}(\vec{r}).
\end{align}
In the above derivation we have used that 
$\left(F^{\alpha}_{\beta}\right)^* = (F^T)^{\beta}_{\alpha}$. Recalling that $(F^{x,z})^T = F^{x,z}$
and $\left( F^y \right)^T = -F^y$ and $\Lambda F^{x,z} \Lambda^T  = -F^{x,z}$ and $\Lambda F^y \Lambda^T = F^y$, which  implies that
\begin{equation}
\mathcal{T} \vec{\hat{F}}  \mathcal{T}^{-1} = 
-\int d\vec{r}\,
c^{\dag}_{\alpha}(\vec{r})  \vec{F}^{\alpha}_{\beta} c^{\beta}(\vec{r}) = -\vec{\hat{F}},
\end{equation}
as expected. We can also apply the same steps to the second quantized form of the 
generators of the SU$(N)$ Lie algebra, which yields
\begin{equation}
\hat{T}^a = \int d\vec{r} \, c^{\dag}_{\alpha}(\vec{r})  \left( T^a \right)^{\alpha}_{\beta} c^{\beta}(\vec{r}).
\end{equation}
which transforms as:
\begin{equation}
\mathcal{T} \hat{T}^a \mathcal{T}^{-1} = \int d\vec{r} \, c^{\dag}_{\gamma}(\vec{r}) \left[ \Lambda^T (T^a)^{T} \Lambda \right]^{\gamma}_{\delta}  c_{\delta}(\vec{r}).
\label{eq:trsta}
\end{equation}
Hence, its thermal average for a Hamiltonian
with TRS (i.e. $\mathcal{T}H\mathcal{T}^{-1} = H$, $\mathcal{T}\rho \mathcal{T}^{-1} = Z^{-1}\, e^{-\beta_T(\mathcal{T} H \mathcal{T}^{-1} - N)} = \rho$) reads
\begin{align}
T^a = \langle \hat{T}^a \rangle &=  \mathrm{Tr}\, \rho\:
\hat{T}^a = \mathrm{Tr}\, \left( \mathcal{T} \rho \mathcal{T}^{-1} \right) \left( \mathcal{T} \hat{T}^a \mathcal{T}^{-1} \right) \notag\\
&= \langle \mathcal{T}^{-1} \hat{T}^a \mathcal{T}^{-1} \rangle = 
\Lambda^T (T^a)^T \Lambda,
\end{align}
where the operator $\mathcal{T}^{-1} \hat{T}^a \mathcal{T}^{-1}$ is given by \eqref{eq:trsta}.
Thus, the order parameter matrix transforms as
\begin{align}
\bar{M} &= \sum_{a=1}^{N^2-1} m_a \langle \hat{T}^a \rangle \\
&= \sum_{a=1}^{N^2-1} m_a \langle \mathcal{T}\hat{T}^a \mathcal{T}^{-1} \rangle  = \Lambda^T \bar{M}^T \Lambda.
\end{align}
Here we have defined the order parameter matrix in terms of the constituent fields $c^{\alpha}(\vec{r}),c^{\dag}_{\alpha}(\vec{r})$ rather than in terms of the Landau quasi-particle density matrix. However, we must recall that quasi-particles carry the same SU($N$) quantum numbers as constituent particles and therefore these two definitions are equivalent. 

\section{Second order free energy}\label{app:pert}

 In this section we provide  derivation of the expression of the (grand canonical) free energy to second order in the interaction strength. The free energy can be obtained from the following expression:
\begin{equation}
F=- \beta^{-1}_T \log \: Z,\label{eq:fnrg}
\end{equation}
where $\beta^{-1}_T = k_B T$ is the absolute temperature and $Z= \text{Tr}\left[e^{-\beta_T H}\right]$ is the partition function in the grand canonical ensemble. For a Fermi system, the latter can be expressed in terms of a Grassmanian functional integral~\cite{Negele1998Quantum}:
\begin{equation}
Z =\int D[\bar{\Psi}, \Psi]\, e^{ -\left(S_0+S_{\text{int}}\right) }.
\end{equation}
In the above expression,  we have introduced
\begin{align}
S_0 &=\frac{1}{\beta}\int^{\beta_T}_0 d\tau \: \sum_{\vec{k},\sigma}\Psi_{\vec{k},\sigma}^*(\partial_{\tau}-\epsilon_\vec{k}+\mu_{\sigma})\Psi_{\sigma}, \\
S_{\text{int}}&= \frac{g}{\beta \Omega}\int^{\beta}_0 d\tau \sum_{\vec{pkq}}\Psi^*_{\vec{p},\alpha}\Psi^*_{\vec{k},\beta}\Psi_{\vec{k-q},\beta}\Psi_{\vec{p+q},\alpha}.
\end{align}
where $\Omega$ is the volume of the system.
Expanding the exponential inside the integral in powers of the interaction action $S_{\text{int}}$ yields the following formal perturbative series for the partition function:
\begin{align}
Z= Z_0 \left( 1-\langle S_{\text{int}}\rangle_0+\frac{\langle \left( S_{\text{int}} \right)^2\rangle_0}{2}+\cdots \right),
\end{align} 
Hence, using Eq.~\eqref{eq:fnrg}, we obtain the perturbative expansion of the free-energy:
\begin{align}
F &= F_0 -\frac{1}{\beta_T}\log \left[1-\langle S_{\text{int}}\rangle^c_0+\frac{\langle \left(S_{\text{int}}\right)^2\rangle^c_0}{2}+\cdots\right],\notag\\
   &=F_0+F_1+F_2+\cdots,
\end{align}
where $\langle\cdots\rangle^c_0$ denotes the sum over connected Feynman diagrams only that result from applying Wick's theorem. The non-interacting free energy is given by the expression: 
\begin{equation}
F_0 = -\frac{1}{\beta_T}\sum_{\vec{k},\sigma}\log[1+e^{-\beta(\epsilon_{\vec{k}\sigma}-\mu_\sigma)}],
\end{equation}
Note that $F_0= \mathcal{E}_0 - T \mathcal{S}_0$, where $\mathcal{E}_0 = \sum_{\vec{k},\alpha} \left( \epsilon_{\vec{k}} - \mu_{\alpha} \right)  n_{\vec{k},\alpha}$
is the internal energy being $n_{\vec{k},\alpha} = \left[e^{\beta_T (\epsilon_{\vec{k}}-\mu_{\sigma})} +1\right]^{-1}$ the Fermi-Dirac distribution function,  and $\mathcal{S}_0 = - k_B \sum_{\vec{k},\sigma} \left[ n_{\vec{k}\sigma} \log n_{\vec{k}\sigma} + \left(1- n_{\vec{k}\sigma}\right) \log \left(1- n_{\vec{k}\sigma} \right) \right]$ is the free gas entropy.

The  leading order corrections  to the non-interacting free energy are:
\begin{align}\label{eq:F1a}
F_1 &= \frac{g}{2\Omega}\sum_{ \alpha\neq\beta}\sum_{\vec{q k}} n_{\vec{q}\alpha} n_{\vec{k}\beta},\\
F_2 &= \frac{  g^2}{ 2 \Omega^2}  \sum_{ \alpha\neq\beta} \sum_{ \vec{pkq} } \frac{ n_{\vec{p}\alpha} n_{\vec{k}\beta}(1-n_{\vec{p+q} \alpha})(1-n_{\vec{k-q} \beta}) }{\epsilon_\vec{p}+\epsilon_\vec{k}-\epsilon_{\vec{p+q}}-\epsilon_{\vec{k-q}}}.
\label{eq:F2a}
\end{align}
Notice that the second order term needs renormalization since the pseudo-potential interaction $\propto g$ has no characteristic momentum cut-off scale and therefore the integral is divergent~\cite{Abrikosov1965,Parthia2011,Huang_Yang_PhysRev.105.767,Huang_PhysRevA.99.063612_2019,Viverit_PhysRevA.69.013607_2004,TAN_AnnPhys_2008}. The renormalization can be carried out by equating the total energy shift of the 2-body system to the physical scattering amplitude, which is proportional to the scattering length $a_s$.
To second order,
\begin{align}
  E^{\text{int}}_{\text{2-body}}&=  \frac{4\pi a_s}{m\Omega}  \notag \\
&\simeq \frac{g}{\Omega} + 
\frac{g^2}{\Omega^2} \sum_{\vec{pkq}} \frac{1}{\epsilon_{\vec{p}}+\epsilon_{\vec{k}}-\epsilon_{\vec{p+q}}-\epsilon_{\vec{k-q}}} 
\end{align}
Inverting the series yields the  following relationship between $g$ and $a_s$ to second order:
\begin{align}
g= \frac{4\pi a_s}{m}- \frac{1}{\Omega}\left(\frac{4\pi a_s}{m}\right)^2\sum_{\vec{pkq}}\frac{1}{\epsilon_{\vec{p}}+\epsilon_{\vec{k}}-\epsilon_{\vec{p+q}}-\epsilon_{\vec{k-q}}},
\end{align}
which can be inserted back into Eq.~\eqref{eq:F1a} and \eqref{eq:F2a},
and leads to the following expression for the renormalized free energyup to second order in $a_s$:
\begin{align}
F&=F_0 + \frac{2\pi a_s}{m\Omega} \sum_{ \alpha\neq\beta}\sum_{\vec{q k}} n_{\vec{q}\alpha} n_{\vec{k}\beta} ,\notag\\
&- \frac{1}{2}\left (\frac{4\pi a_s}{m\Omega}\right)^2  \sum_{ \alpha\neq\beta} \sum_{ \vec{pkq} } \frac{ n_{\vec{p}\alpha}  n_{\vec{k}\beta} ( n_{\vec{p+q} \alpha} + n_{\vec{k-q} \beta} ) }{\epsilon_{\vec{p}}+\epsilon_{\vec{k}}-\epsilon_{\vec{p+q}}-\epsilon_{\vec{k-q}}}.
\end{align}
In the main text this expression is minimized by treating the  $\mu_{\sigma}$ as variational parameters. This method improves  the results of  the unrestricted Hartree-Fock method  while capturing some of the effects of fluctuations beyond mean-field theory.

\section{Numerical minimization method}

In order to find the global minimum of the free energy, we start from an initial choice of parameters (i.e. chemical potentials $\mu^{(0)}_1,\cdots,\mu^{(0)}_N$), which determine the population of each species that allow us to carry out the minimization using the Basic Differential Multiplier Method (BDMM~\cite{Platt_NIPS_1987}).  The integral of the second order correction to the free-energy is computed  using the  method described in~\cite{Conduit_PhysRevA.79.053606_2009}. Following this method, the integral:
\begin{align}
\int &F(|\vec{k_1}|,|\vec{k_2}|,|\vec{k_3}|,|\vec{k_4}|) \\& \delta(\vec{k_1}+\vec{k_2}-\vec{k_3}-\vec{k_4})\notag
 d\vec{k_1}d\vec{k_2}d\vec{k_3}d\vec{k_4},
\end{align}
is reduced to the following four-dimensional integral:
\begin{align}
16\pi^3\int &F(k_1,k_2,k_3,k_4)k_1k_2k_3k_4\notag\\
&\times \text{max}\biggl[ 0,\text{min}(k_1+k_2,k_3+k_4)\notag\\
&-\text{max}(|k_1-k_2|,|k_3-k_4|)\biggr] dk_1 dk_2 dk_3 dk_4,
\end{align}
which is evaluated numerically. Using this result, besides the second order correction to the free
energy, we also compute and its derivatives with respect to the chemical potentials $\mu_{\alpha}$  The latter are required to find the global minimu of the free-energy.

In order to impose the  constraint of  constant total particle density in the BDMM,  we consider the following generalization of the free energy:
\begin{align}
F^{\prime}(\mu_1,&...,\mu_N,T ;\lambda)=F(\mu_1,\cdots,\mu_N,T)
,\notag\\ 
&+\lambda\left[ n_1(\mu_1,T)+\cdots+n_N(\mu_N,T) - n_{\text{tot}}\right], \notag \\
&+b\left[ n_1(\mu_1,T)+\cdots+n_N(\mu_N,T)- n_{\text{tot}}\right] ^2,
\end{align}
where  $n_{\alpha}(\mu_{\alpha},t)$ is the particle density for species $\alpha$. In the above generalized free energy,  $\lambda$ is an additional Lagrange multiplier  introduced to numerically enforcethe constant-density constraint. The additional square term proportional  to the constant $b$ is used to enhance the stability and convergence towards the minimum. The optimization is carried out by updating $\mu_1,\cdots,\mu_N$ and $\lambda$ in small steps, $s_\mu$ and $s_\lambda$~
\footnote{Technically, we chose $s_\lambda=10 s_\mu$ to ensure the convergence on total particle number is faster than the optimization of $\mu_i$, in order to prevent the optimization from jumping to another local minimum of $\mu_i$} according to 

\begin{align}
\mu_{N}^{ (n+1) } &= \mu_{N}^{ (n) }  -  s_\mu \frac{ \partial F' ( \mu^{(n)}_1,\cdots,\mu^{(n)}_N,T;\lambda^{(n)} ) }{\partial \mu_{N}^{(n)} },\notag\\
\lambda^{ (n+1) } &= \lambda^{ (n) }  + s_{\lambda} \frac{ \partial F' ( \mu^{(n)}_1,\cdots,\mu^{(n)}_N,T;\lambda^{(n)} ) }{\partial \lambda^{(n)} }.
\end{align}
$\mu_N^{(0)}$ and $\lambda^{(0)}$ are determined by finding the smallest energy of the sampling points by direct computation. The minimization is carried out by minimizing the free energy $F^{\prime}$ with respect to $\mu_1,...\mu_N$ and maximizing it with respect to $\lambda$. 
\section{Minimization using Kanno's formula}
\label{app:kanno}
\begin{figure}[h]
\center
\includegraphics[width=\columnwidth]{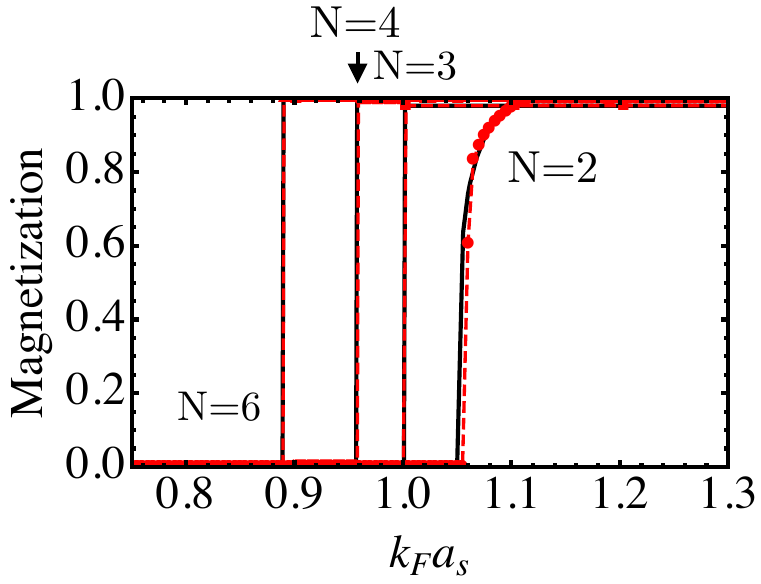}
\caption{ Magnetization at zero temperature derived from  numerical integration of the second order free energy (red dashed curves) and using Kanno's formula~\cite{Kanno_PTP_1970} (solid black curves). In both cases, we perform an unconstrained numerical minimization of the total  energy. From right to left, the curves correspond to $N = 2, 3, 4, 6$, respectively for a three-dimensional Fermi gas with SU$(N)$ symmetry.}
 \label{fig:Kanno}
\end{figure}
The free energy up to second order in the gas parameter  can be written as,
 \begin{align}
     F=\sum_{\sigma} f_{0;\sigma}+ \sum_{\alpha\beta;\alpha\neq\beta} f_{1;\alpha\beta}+\sum_{\alpha\beta;\alpha\neq\beta}f_{2;\alpha\beta},
 \end{align}
where $f_0$ and $f_1$ can be easily parameterized by the density of  each component allowing an analytical expression of the free energy at mean-field level,
 \begin{align}
     f_{0;\sigma}=\frac{3n_{\text{tot}}k_F^2}{10 m}\frac{p_\sigma^5}{2}
 \end{align}
\begin{align}
     f_{1}(p_\alpha,p_\beta)=\frac{(k_Fa_s)n_{\text{tot}}k_F^2}{3\pi m}p_\alpha p_\beta
\end{align}
where $p_{\sigma}= (N n_{\sigma})^{1/3}$. For the second order correction, 
at zero temperature, we can directly use the result of Kanno\cite{Kanno_PTP_1970} which yields the following expression,
\begin{align}
    f_2(p_\alpha,p_\beta)=\frac{3 n_{\text{tot}}k_F^2}{10 m}\left[ \left(\frac{k_Fa_s}{\pi}\right)^2 I(p_\alpha,p_\beta)\right],
\end{align}
where
\begin{align}
I(p_{\alpha},p_{\beta})&=\frac{1}{21}\biggl[
     22p{_\alpha}^3 p_{\beta}^3 (p_{\alpha}+p_{\beta})\notag\\
&-4p_{\alpha}^7\log\left(\frac{p_{\alpha}+p_{\beta}}{p_{\alpha}}\right)-4p_{\beta}^7\log\left(\frac{p_{\alpha}+p_{\beta}}{p_{\beta}}\right)\notag\\
     &+\frac{1}{2}(p_{\alpha}-p_{\beta})^2 p_{\alpha}p_{\beta}(p_{\alpha}+p_{\beta})\notag\\
&\times\left[15(p_\alpha^2+p_{\beta}^2)+11p_{\alpha}p_{\beta}\right]\notag \\
     &-\frac{7}{8}(p_\alpha-p_{\beta})^4(p_{\alpha}+p_{\beta})\notag\\
    &\times\left[ (p_{\alpha}+p_{\beta})^2+p_{\alpha}p_{\beta}\right]\log\left(\frac{p_{\alpha}+p_\beta}{|p_\alpha-p_\beta|+0^+}\right)^2
     \biggr].
 \end{align}
 Minimizing the free energy with the constraint $\sum_\sigma n_\sigma=n_\text{tot}$ allows us to obtain the equilibrium magnetization. 
 Fig.~\ref{fig:Kanno} shows the magnetization at zero temperature ($T=0$) obtained using Kanno's integral and the numerical integration of the 2nd
 order free-energy expression at $T=0$. The agreement between two approaches is excellent.
 
\bibliography{ferro}

\end{document}